\documentclass{revtex4}

\usepackage{amsmath}
\usepackage{graphicx}
\usepackage{subfig}

\DeclareMathOperator{\sech}{sech}

\DeclareMathOperator{\arccosh}{arccosh}

\usepackage{array}
\def\ba{\begin{eqnarray}}
\def\ea{\end{eqnarray}}
\def\beq{\begin{equation}}
\def\eeq{\end{equation}}

\begin{document}

\title{Tearing of Free-Standing Graphene}
\author{M. J. B. Moura}
\affiliation{IBM Research - Brazil, \\
Av. Pasteur 138/146, Botafogo, Rio de Janeiro, Brazil, CEP 22290-240}
\author{M. Marder}
\email{marder@mail.utexas.edu: To whom correspondence should be addressed.}
\affiliation{Center for Nonlinear Dynamics and Department of Physics\\
1 University Station C1600, Austin, TX 78712-0264.}

\maketitle

\section*{Abstract}

We examine the fracture mechanics of tearing graphene.  We present a
molecular dynamics simulation of the propagation of cracks in clamped,
free-standing graphene as a function of the out-of-plane force.  The
geometry is motivated by experimental configurations that expose
graphene sheets to out-of-plane forces, such as back-gate voltage.  We
establish the geometry and basic energetics of failure, and obtain
approximate analytical expressions for critical crack lengths and
forces.  We also propose a method to obtain graphene's toughness.  We
observe that the cracks' path and the edge structure produced are
dependent on the initial crack length.  This work may help avoid the
tearing of graphene sheets and aid the production of samples with
specific edge structures.

\section{Introduction}
\label{sec_intro}

Fracture mechanics provides an ingenious framework by which to
assemble experimental information and calculations from continuum
elasticity into a theory of material failure. This theory is
particularly important for structures that are subject to many design
constraints but must be protected against failure at all costs; for
example, aircraft \cite{Broek.94}.

Graphene is a single layer two-dimensional honeycomb lattice of carbon
atoms.  It is light, flexible, thermally stable (in a non-oxidizing
environment), and has a high electrical conductivity (the Fermi
velocity of electrons in graphene is $10^6\,\textrm{m/s}$
\cite{Geim}).  Many experiments have used graphene in the
free-standing experimental setup, mostly in an effort to improve
electronic properties through absence of a substrate
\cite{garcia-sanchez_imaging_2008,bolotin_ultrahigh_2008,bolotin_temperature-dependent_2008,zande_large-scale_2010,Dorgan.13} 

Graphene is very stiff against in-plane distortions, with a Young
modulus on the order of $\approx 1\,\textrm{TPa}$
\cite{lee_measurement_2008}.  Because it is so thin, it is very
floppy, and the bending energy is of the order of 1eV
\cite{nicklow_lattice_1972-1}.  The large value of Young's modulus
gives graphene a very high ideal breaking stress of around 130 GPa,
leading to the assertion that it is the strongest material known
\cite{lee_measurement_2008}. This statement is true for defect-free
samples. However in practical applications, sample defects are almost
inevitable on some scale, and therefore the toughness of graphene is
important to determine.

Previous studies of fracture of graphene have employed in-plane
geometry
\cite{terdalkar_nanoscale_2010,kim_ripping_2011,lu_atomistic_2011,lu_effect_2010}.
The energy cracks require to create a new surface in graphene, the
fracture toughness, should not be expected to depend upon the precise
path by which atoms pull apart around the crack tip, except to the
extent that dissipative processes such as generation of phonons is
involved \cite{Holland.99}. Thus in-plane calculations should not be
appreciably worse than the out-of-plane calculations we will perform
to obtain the fracture toughness of graphene. However, when it comes
time to compare with experiment, details of the geometry do
matter. Empirical atomic potentials are not necessarily very reliable
when it comes to the details of surface
energies\cite{Holland.99}. Toughness is better obtained from
experiment than from theory. To make this determination possible, one
needs a geometrical setting where theory and experiment coincide and
fracture toughness is the only unknown parameter.  Then toughness can
be obtained from critical values of force or initial crack length
where cracks begin to run.

In some experimental configurations free-standing graphene samples
show cracks and holes, and sometimes the samples break
\cite{kim_ripping_2011,zande_large-scale_2010,Insun,Bolotin,Goldberg}.
The back-gate voltage experimental setup is an example.  In this case
the free-standing graphene sheet is pinned at its edges and suspended
between two shelves. The graphene experiences a downward force because
the substrate has a voltage different from the graphene itself.  We
perform molecular dynamics simulations in this geometry, employing
nanometer scale samples. From these simulations, we deduce the basic
geometry of the failure process. With the understanding of the
geometry in hand, we develop analytical approximations that can be
employed on samples of any size where failure occurs in the same
mode. We verify that our simulations are in satisfactory accord with
these approximations.

The structure of this article is as follows: In Section
\ref{sec_numerical} we describe our simulations and point out the
essential way that sheets deform in the presence of uniform downward
forcing to drive crack motion. In Section \ref{sec_analytical} we
develop analytical expressions to correspond to this geometry. In
Section \ref{sec_comparison} we compare our analytical expressions
with our simulations and the available experimental evidence to
estimate fracture toughness of graphene. Section \ref{sec_conclusion}
contains final discussion and conclusions.

\section{Numerical study}
\label{sec_numerical}

Here we present a numerical study of the propagation of cracks in
clamped, free-standing graphene as a function of the out-of-plane
force.  We use the MEAM semi-empirical potential
\cite{baskes_modified_1992}, shown to reproduce well the properties of
graphene \cite{thompson-flagg_rippling_2009,lee_modified_2005} and to
support crack propagation \cite{holland_cracks_1999}.  The energy
minimization is done through damped molecular dynamics.  To better
model experimental conditions, all the simulations in this work
are of finite-sized graphene sheets, and no periodic boundary
conditions are used.  The simulations were done at zero temperature so
as to focus on fracture mechanics. We saw no indication during the
investigation of phenomena such as lattice trapping where thermal
fluctuations would have been important to overcome energy barriers.

Cracks on graphene sheets have been observed to come in multiple sizes
and shapes \cite{zande_large-scale_2010}.  Most experiments focus on
electronic properties, and do not look at initial cracks on the
samples.  Consequently, the shapes of the cracks are in general
unknown.  We consider two possible initial conditions: a crack in the
middle of the sheet and a crack at the very edge of the sheet.  The
first because it is a symmetrical problem and the second because
defects can occur where the graphene sheet meets the support that
suspends it.

\begin{figure}[h]
  \subfloat[initial crack of $l \approx$ 10{\AA}]{\label{fig_side_10A_initial}\includegraphics[width=0.33\textwidth]{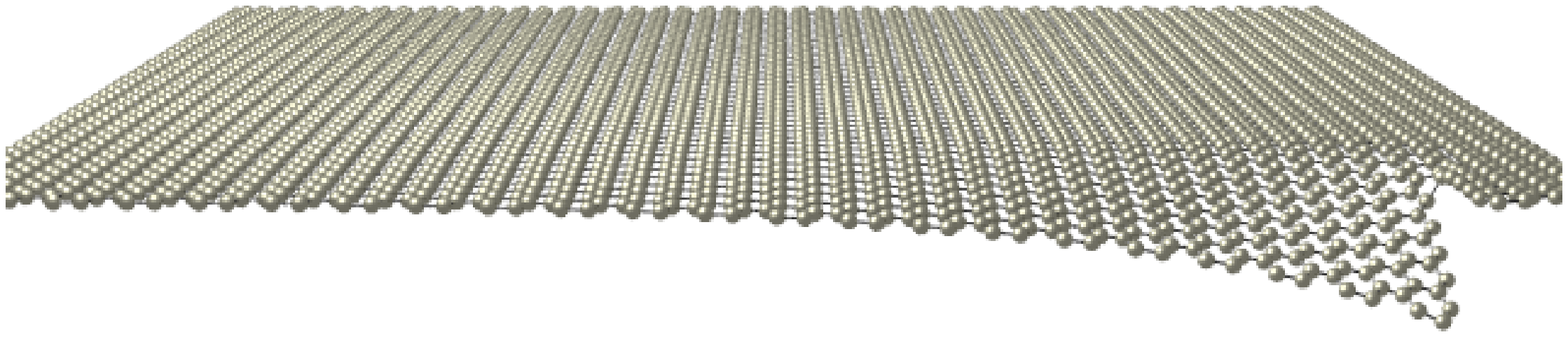}}
  \subfloat[sheet wrinkles and bends]{\label{fig_side_10A_inbetween}\includegraphics[width=0.33\textwidth]{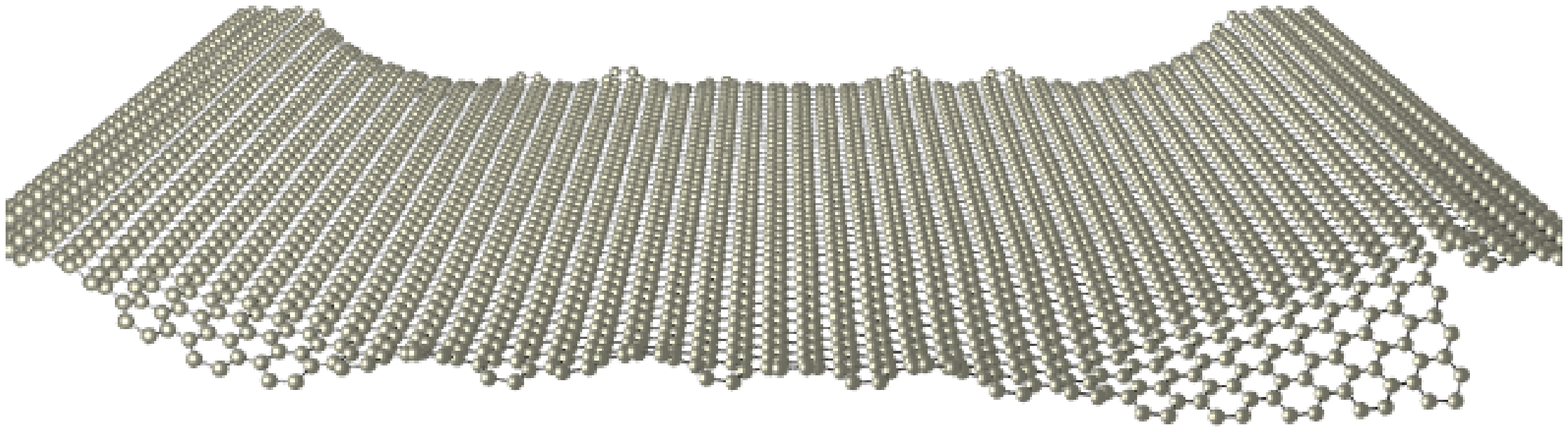}}
  \subfloat[crack propagates]{\label{fig_side_10A_running}\includegraphics[width=0.33\textwidth]{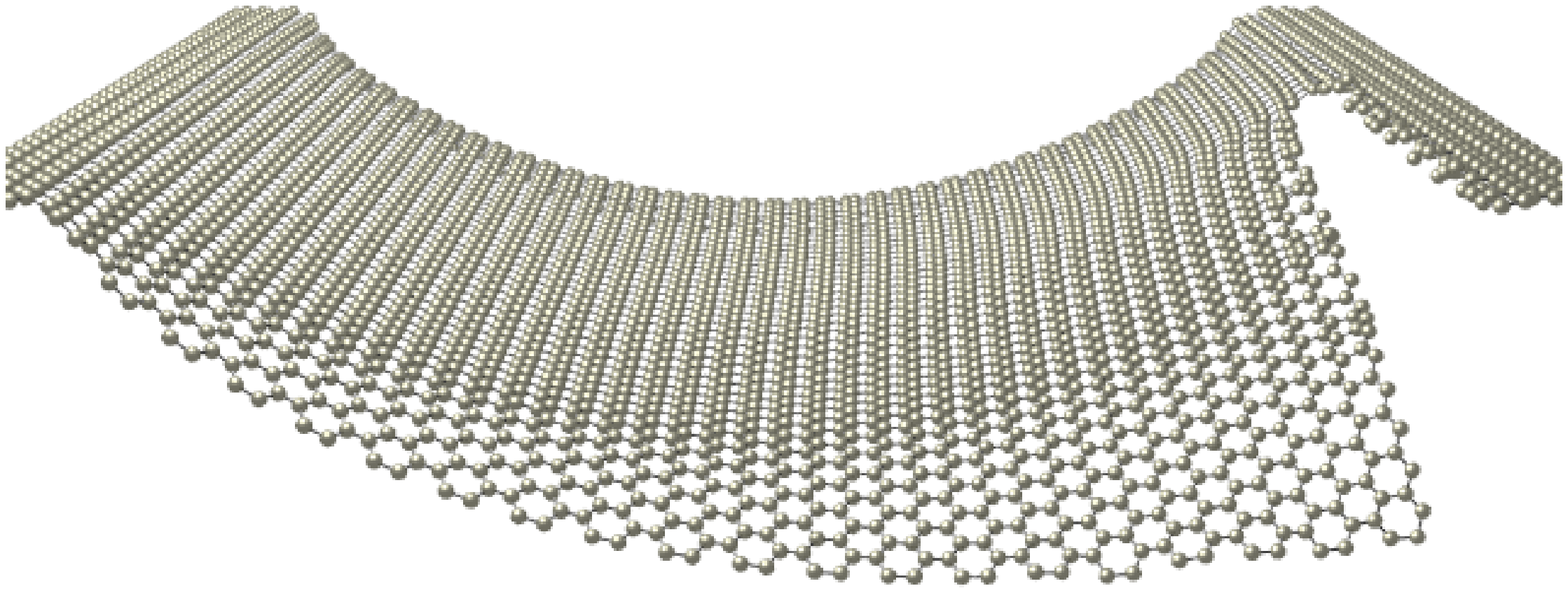}}
\caption{Constant downward force is applied on a 
100{\AA} by 100{\AA} clamped, free-standing graphene sheet
with an initial crack of $l \approx$ 10{\AA}. Notice that the sheet wrinkles and bends before the crack runs.}
  \label{fig_side_crack}
\end{figure}

We start with a flat graphene sheet, where the position of the atom
$i$ is represented by $(x_0^i, y_0^i, z_0^i)$. We consider a straight
crack that runs parallel to the x-axis and has a length of
$x_{cut}$. To provide a seed configuration we keep the $x$ and $y$ position of the
atom and change $z$ by
 \beq z^i = z^i_0 - P_{A} \left(1- \exp{(P_{B}
    \cdot y_0^i)} \right) \cdot \left( 1 - \exp{(P_{B}(-x_0^i +
    x_{cut}))} \right)
\label{numericalbend} 
\eeq 
The parameters $P_{A}$ and $P_{B}$ determine the initial
curvature of the sheet.  As an example of an initial condition,
Fig.~\ref{fig_side_10A_initial} shows a crack of $l \approx$ 10{\AA}
at the edge of a suspended graphene sheet.  The clamped edges are not
allowed to move and where chosen to have a width of $\approx$ 5{\AA}. 

During the initial molecular dynamics time steps the sample relaxes, as
shown in Fig.\ref{fig_side_10A_inbetween}. Given sufficient time in
the absent of external forces, the crack would zip back up and the
sample would heal. This does not happen because we apply a downward
force to every atom in the sheet. For samples where sufficient force is applied,
the crack starts running, as seen in 
Fig.~\ref{fig_side_10A_running}.  Short movies of crack propagation
are available at \cite{videos.12}.

We do not know what creates cracks in graphene sheets in actual
experiments. They result from impurities, defects, or details of
sample preparation.  We emphasize that in the theory of fracture
mechanics, crack propagation is independent of the mechanism that
initially creates the crack.

\begin{figure}
\includegraphics[width=.5\columnwidth]{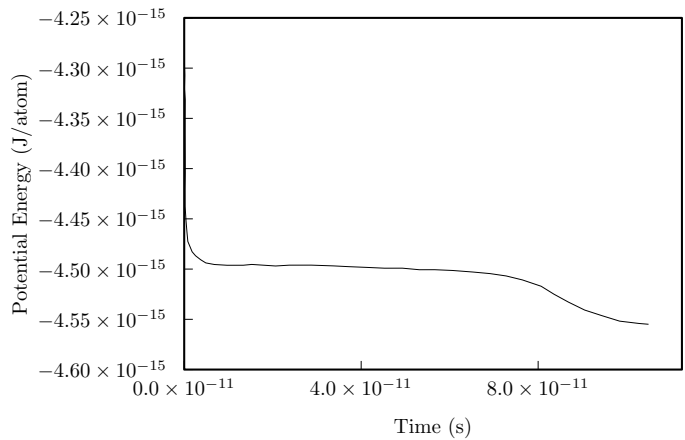}
\caption{Potential energy  vs time for an initial edge crack of $l \approx$ 35{\AA}. 
The first large drop in potential energy happens while the sheet ripples and bends.
Then potential energy slowly decreases as the ripples disappear and the sheet reaches its fully bent state. 
Another fast drop in potential energy occurs when the crack starts running.} 
\label{fig_side_crack_pe}
\end{figure}

We encountered some difficulties in determining the period of time
that it takes for a crack to begin to run.  The simulations show that
first the sheet ripples and bends, and then the crack runs.  In terms
of energy, what we see is an initial large drop in potential and
kinetic energy, Fig.~\ref{fig_side_crack_pe}.  Then the sheet reaches
an almost-stable state, where the energy almost plateaus, decreasing
very slowly.  Finally, when the crack runs, another drop in potential
energy occurs together with a fast increase in kinetic energy. If the
force applied is not strong enough for the crack to run, the sheet
stays in the bent state forever. Numerically we have to set an
acceptable period of time to be considered ``forever''.  We observe
that longer initial crack lengths lead to longer periods of time spent
in the almost-stable state.  After studying many simulations we
decided that 600,000 time steps $=1.5\times 10^{-10}$ s is, in most
cases, an acceptable period of time to study the crack propagation (or
lack of it). 

\begin{figure}[h]
  \subfloat[initial edge crack of $l \approx$ 20{\AA}]{\label{fig_edge_crease}\includegraphics[width=0.33\textwidth]{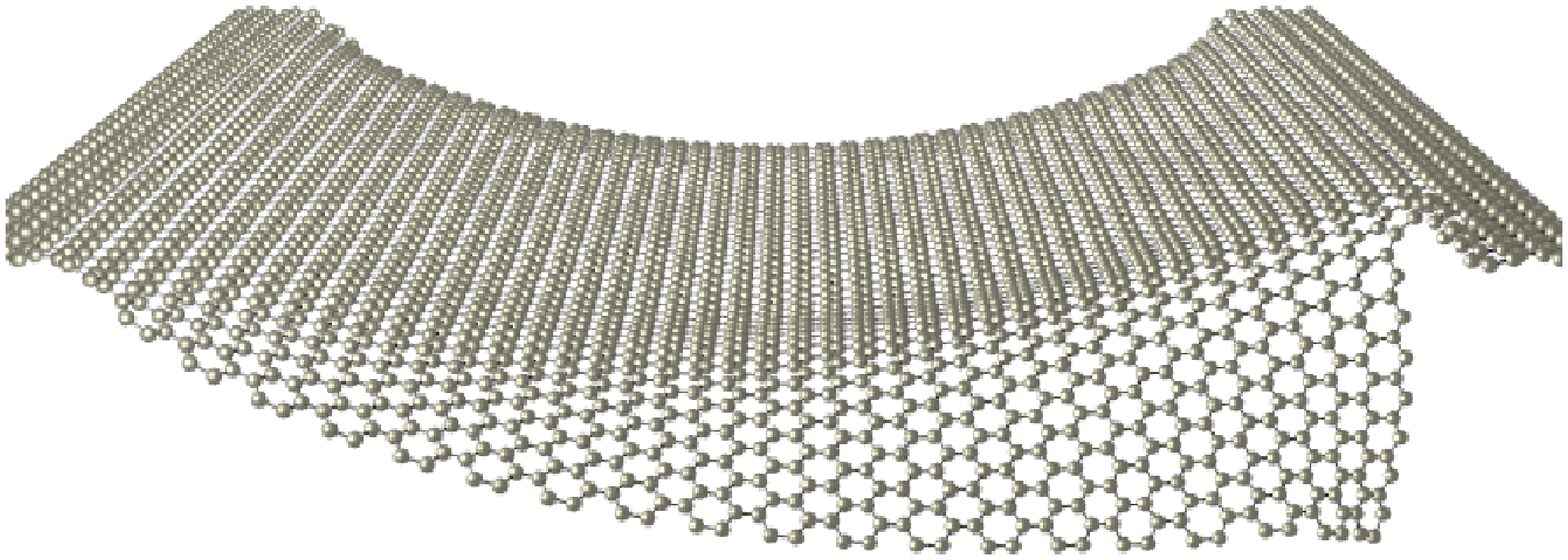}}
  \subfloat[initial middle crack of $l \approx$ 25{\AA}]{\label{fig_middle_crease}\includegraphics[width=0.33\textwidth]{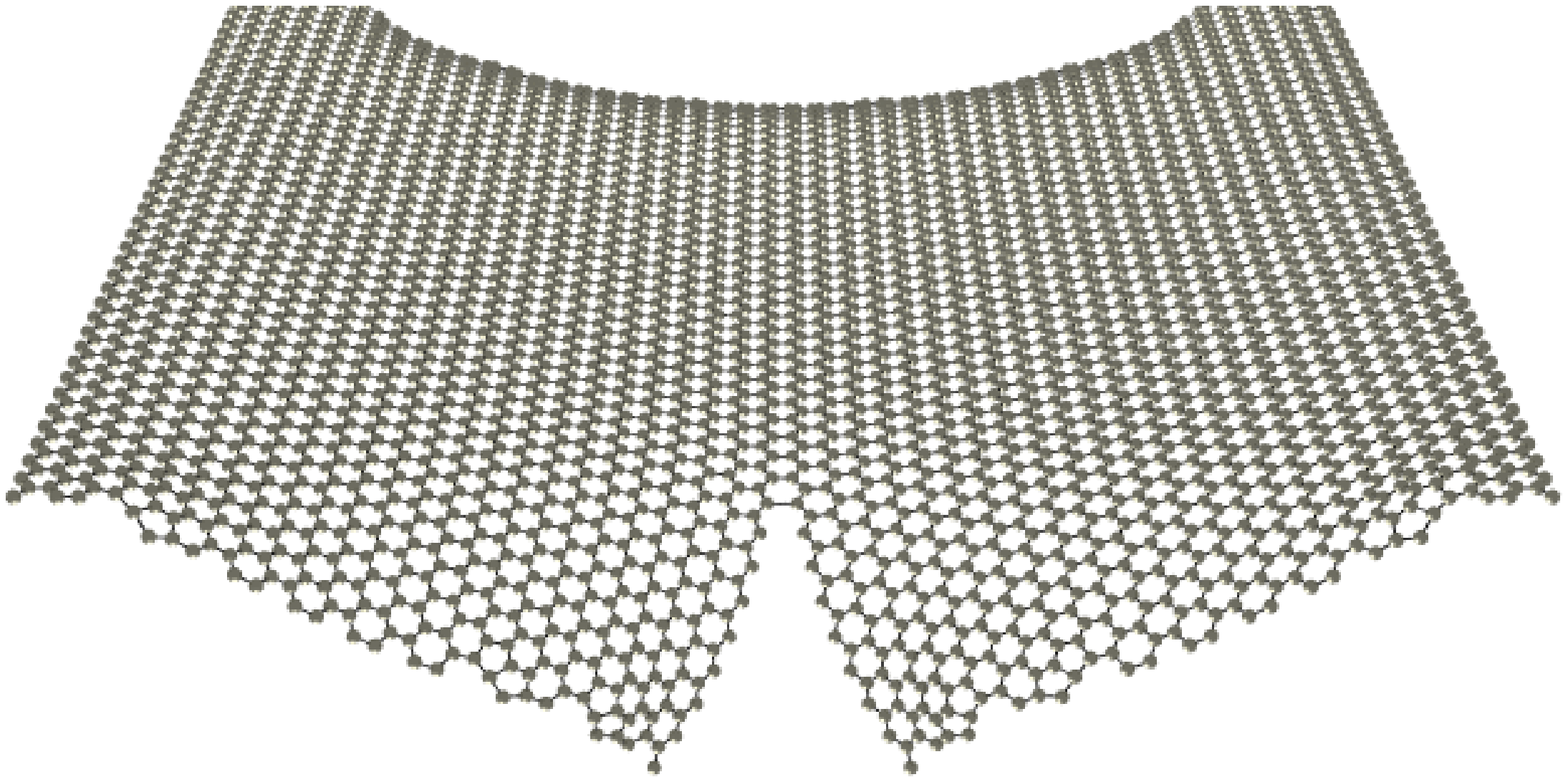}}
\caption{Clamped, free-standing graphene sheets with an initial crack (at
the edge and in the middle) under a downward constant force. In both cases
the sheet exhibits a crease before the crack runs. The crease goes from the
crack tip essentially all the way to the fixed end.}
  \label{fig_crease}
\end{figure}

The numerical simulations show the basic geometry of the failure process. 
In both cases, cracks at the edge and in the middle of the sheet, we observe that the
sheet folds in a crease (on both sides of the crack) before the crack runs, Fig.~\ref{fig_crease}. In the next section we use this geometry to 
obtain approximate analytical expressions for critical crack lengths and forces.

%

\section{Analytical approach to tearing a two dimensional sheet}
\label{sec_analytical}

The system of interest is a two dimensional sheet, such as graphene, with an initial crack of length $l$. 
The sheet is suspended and 
exposed to a uniform downward force $f$. 
The problem is to describe the propagation of a crack in such sheet and the minimum force required for the crack 
to run.

Here we follow a procedure similar to the one developed by Marder \cite{marder_effects_2004} 
for the propagation of a crack in a 3D strip, making the appropriate changes for our two dimensional problem.

\begin{figure}
\includegraphics[width=0.5\columnwidth]{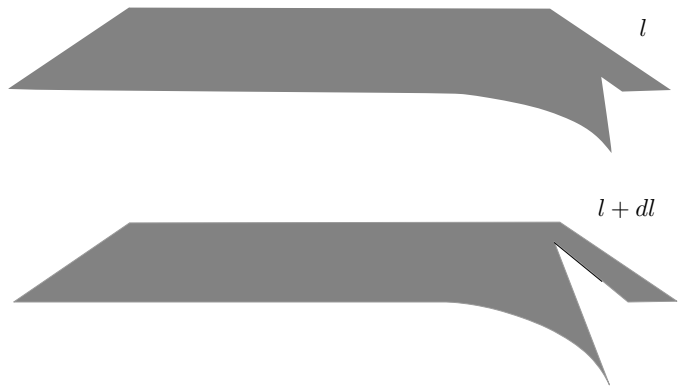}
\caption{Top image: Clamped, free-standing two dimensional sheet with an initial crack of length $l$ at the right hand side. 
The sheet is clamped at the left and right edges. Bottom image: Because of an external downward force the crack runs, a length $dl$, and the sheet bends diagonally.}
\label{fig_2Dcrackrun}
\end{figure}

Consider a system with an initial crack of length $l$ and total energy $U_{tot}(l)$, 
Fig.~\ref{fig_2Dcrackrun}.
The crack can run a length $dl$ if doing so reduces the total energy of the system; that is:
\beq
 U_{tot}(l)>U_{tot}(l+dl)
\label{minenergy}
\eeq

The total energy of the system can be written as the energy contained within the crack tip region plus the energy outside of it, 
$U_{out}$. 
The energy to move the crack tip (region) is proportional to the energy of the new surface opened up by the crack. Therefore the total 
energy of a 2D sheet, such as graphene, with a crack of length $l$ is given by:
\beq
 U_{tot}(l) = \Gamma l + U_{out}(l)
\label{utot}
\eeq
where $\Gamma$ is the fracture toughness. The fracture toughness is material-dependent, and it can be measured experimentally 
(and obtained numerically).

From Eqs.~(\ref{minenergy}) and (\ref{utot}) we get Griffith's criterion for a crack to propagate in a 2D sheet:
\beq
 \frac{dU_{out}}{dl} + \Gamma < 0.
\label{griffiths}
\eeq 
This should be understood as a necessary but not sufficient
condition for crack motion. Atomic systems can exhibit {\sl lattice
  trapping}, where cracks become stuck between atoms and do not
propagate even though there is enough energy available to allow it
\cite{Hsieh.73}. This phenomenon leads to hysteresis depending upon
whether the driving force for crack motion is increasing or
decreasing. As we ramped external forces both up and down and saw
cracks run forward or heal backward at very nearly the same force,
provided we waited long enough, we do not think lattice trapping is
very important in this case. We analyze the fracture mechanics of the two geometries under consideration: 
a crack at the very edge of the sheet and a crack in the middle of the sheet.

\subsection{Analytical study of a sheet with an edge crack}
\label{sec_theoryedge}

For an initial crack at the edge of the sheet the downward force will tear the sheet at the edge making it 
bend diagonally, as seen on Fig.~\ref{fig_2Dcrackrun}. 
The energy outside the crack tip, $U_{out}$, is then equivalent to the energy required to bend a 2D sheet.

\begin{figure}
\includegraphics[width=7cm]{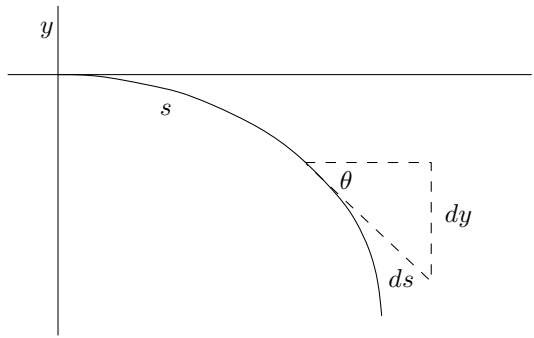}
\caption{One dimensional strip of length $L$, fixed at one end, and bending
due to an external downward force.}
\label{fig_1Dbending}
\end{figure}

First, we consider the energy needed to bend a strip (Fig.~\ref{fig_1Dbending}). 
Then we extend the result to the 2D case of a bending sheet.

%

\subsubsection{Energy for a bending strip}
\label{sec_1Dbending}

The energy of a 1D strip of length $L$ under a downward force $f$ is given by:
\begin{equation}
U^{1D} = \int_{0}^{L} ds \left[\frac{k_l}{2} \left(\frac{d \theta}{ds}\right)^2 + f_{l} \cdot y(s)\right]
\label{U1Dbend}
\end{equation}
where $k_l$ is the bending modulus (times length). The $\theta$ term refers to the bending energy 
and the $f_{l}$ term refers to energy due to the external downward force (per length) applied to the strip. 
From Fig.~\ref{fig_1Dbending} we observe that:
\begin{equation}
 y(s)=\int \sin(\theta) ds
\label{no_small_bend}
\end{equation}
Therefore: 
\begin{equation}
U^{1D} = \int_{0}^{L} ds \left[\frac{k_l}{2} \left(\frac{d \theta}{ds}\right)^2 + f_{l} \int_{0}^{s} \sin(\theta(s')) ds'\right]
\label{U1D_no_small_bend}
\end{equation}

To obtain the energy outside the crack tip in terms of the minimum force required for the crack to run 
we need to minimize Eq.~(\ref{U1D_no_small_bend}), which results in (see Appendix \ref{sec_full_1Dbending} for details):
\beq
U^{1D}= 2\left(2- \sqrt{2}\right)\sqrt{f_l k_l L} -\frac{1}{2}f_lL^2
\label{U1D_crease_large_L}
\eeq
The first term refers to the folding of the strip and the second term
to the potential energy.  We note that the folding energy depends,
somewhat surprisingly, on the total length $L$ of the strip. This
happens because as the strip length increases so does the force on it,
and the crease bends at a tighter and tighter angle.

\subsubsection{Energy for a bending sheet}
\label{sec_2Dbending}

\begin{figure}
\includegraphics[width=4cm]{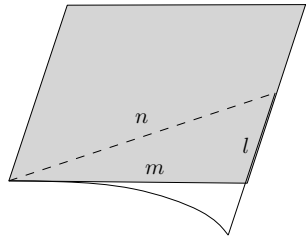}
\caption{Suspended sheet with an initial crack of length $l$. 
The triangle formed by the crease is the part of the sheet that is initially free to bend, 
as it is not attached to the support. 
}
\label{fig_2Dcrease}
\end{figure}

Now we obtain approximate expressions for a sheet folding from the crack tip all the way to the fixed end, 
forming a crease as seen in Fig.~\ref{fig_2Dcrease}. 
We consider the bending energy for a right triangle, with one side of length $l$, 
the crack length, and another side of length $m$, the horizontal width of the crease. 

%
\begin{figure}
\includegraphics[width=4cm]{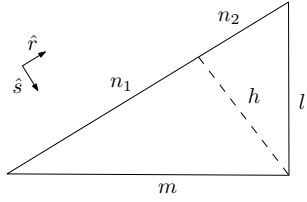}
\caption{Relabeling Fig.\ref{fig_2Dcrease}: right triangle formed by
  the crease, $n=n_1+n_2$, the crack length, $l$,  
and the horizontal width of the crease, $m$. 
Note that the horizontal width of the crease, $m$, is essentially the width of the sheet.
}
\label{fig_triangle_crease}
\end{figure}

  In the previous section we obtained the energy of a bending strip under a downward force. We can now integrate that result over the right triangle 
to obtain the bending energy of a triangle. 
We break the triangle into two parts (Fig.~\ref{fig_triangle_crease}) and obtain:
\beq
U^{2D}=U_1 + U_2  
\label{U1U2_crease}
\eeq
where:
\beq
U_1 = \int_{0}^{n_1} dr_1 \int_{0}^{L(r_1)} ds \left[\frac{k}{2} \left(\frac{d \theta}{ds}\right)^2 + 
f_{a} \int_{0}^{s} \sin(\theta(s')) ds'\right]
\label{U1}
\eeq
and 
\beq
U_2 = \int_{0}^{n_2} dr_2 \int_{0}^{L(r_2)} ds \left[\frac{k}{2} \left(\frac{d \theta}{ds}\right)^2 + 
f_{a} \int_{0}^{s} \sin(\theta(s')) ds'\right] .
\label{U2}
\eeq
Here $f_{a}$ is the downward force per area and $k$ is the bending
modulus (units of energy).

Notice that, in the two-dimensional case, the total bending length, $L(r)$, 
varies throughout the triangle, and is written as $L(r_1)=({h}/{n_1})r_1$ 
for the first triangle, and $L(r_2)=({h}/{n_2})r_2$ for the second triangle. 
The height of the triangle is given by $h={m \cdot l}/{\sqrt{m^2 + l^2}}$ and the hypotenuse by $n=n_1+n_2=\sqrt{m^2+l^2}$ 
(see Fig.~\ref{fig_triangle_crease}).

The first integral is the same as the one we solved for the 1D case of the bending strip. Therefore, 
in the case of $U_1$ for example:
\beq
U_1 = \int_{0}^{n_1} dr_1 
\left[
2 \left(2- \sqrt{2}\right)\sqrt{k f_a L(r_1)} -\frac{1}{2}f_aL^2(r_1)
\right]
\eeq

Substituting $L(r_1)=hr_1/n_1$ and solving the integral in $r_1$ we obtain:
\beq
U_1 = 
n_1\left[ 
\frac{4}{3} \left(2- \sqrt{2}\right)\sqrt{k f_a h}
-\frac{1}{6}f_ah^2
\right]
\eeq

As $U_2$ is analogous to $U_1$, Eq.~(\ref{U1U2_crease}) results in:
\beq
U^{2D}=U_1 + U_2 = (n_1+n_2) \left[ 
\frac{4}{3} \left(2- \sqrt{2}\right)\sqrt{k f_a h}
-\frac{1}{6}f_ah^2\right]
\eeq

Substituting $n=n_1+n_2=\sqrt{m^2+l^2}$ and $h=\frac{ml}{\sqrt{m^2 + l^2}}$, we obtain that 
the energy required to bend a right triangle is given by:
\beq
U^{2D}=  C\sqrt{k f_a n m l } - \frac{1}{6}f_a\frac{m^2l^2}{n}
\label{U2D_crease}
\eeq
where $C=\frac{4}{3} \left(2- \sqrt{2}\right)$,
$f_{a}$ is the downward force per area, $k$ is the bending modulus, $m$ is the horizontal width of the crease, 
$l$ is the crack length, and $n=\sqrt{m^2+l^2}$.
Note that, the first term refers to the folding of the sheet and the second term to the potential energy.

\subsection{Griffith Point}
\label{sec_griffiths}

In the beginning of Sec.~\ref{sec_analytical} we presented Griffith's criterion for a crack to propagate 
on a 2D sheet:
\begin{equation*}
 \frac{dU_{out}}{dl}+\Gamma=0
\end{equation*}

Substituting the energy outside the crack tip, $U_{out}$, for the energy to bend a triangle $U^{2D}$, Eq.~\eqref{U2D_crease}, 
and solving for the minimum force (per area) for the crack to run, we obtain our {\bf final expression for an edge crack}:
\ba
f_a^{edge}=\frac{3(l^2 + m^2)^{3/2}}{2 l^3 m^3 (l^2 + 2 m^2)^2}\Big[3 C^2 k (2 l^2 + m^2)^2 + 4 \Gamma l^2 m (l^2 + 2 m^2) \nonumber \\ 
  + C (2 l^2 + m^2) \sqrt{3 k (3 C^2 k (2 l^2 + m^2)^2 + 8 \Gamma l^2 m (l^2 + 2 m^2))}\Big]
\label{fgeneral_edge}
\ea
where 
$k$ is the bending modulus \cite{nicklow_lattice_1972,khare_coupled_2007,thompson-flagg_rippling_2009,fasolino_intrinsic_2007}, $\Gamma$ is the fracture toughness, $C=\frac{4}{3} \left(2- \sqrt{2}\right)$, $l$ is the crack length, and $m$ is the horizontal width of the crease. 
See Fig.~\ref{fig_2Dcrease}. Note that the horizontal width of the crease, $m$, is essentially the width of the sheet.

For a crack in the middle of the sheet, as there are two folds, the total energy outside the crack tip is:
\begin{equation}
U_{out} = 2 \cdot U^{2D}
\end{equation}

Hence our {\bf final expression for a crack in the middle of the sheet} is:
\ba
f_a^{middle}=\frac{3(l^2 + \tilde{m}^2)^{3/2}}{2 l^3 \tilde{m}^3 (l^2 + 2 \tilde{m}^2)^2}\Big[3 C^2 k (2 l^2 + \tilde{m}^2)^2 
+ 2 \Gamma l^2 \tilde{m} (l^2 + 2 \tilde{m}^2) \nonumber \\ 
  + C (2 l^2 + \tilde{m}^2) \sqrt{3 k (3 C^2 k (2 l^2 + \tilde{m}^2)^2 + 4 \Gamma l^2 \tilde{m} (l^2 + 2 \tilde{m}^2))}\Big]
\label{fgeneral_middle}
\ea
where $\tilde{m}$ is essentially {\bf half} the width of the sheet for a sheet with crack in the middle.

For initial cracks that are much smaller than the width of the sheet ($l<<m$) we find:
\beq
f_a^{edge_{l<<m}}= 3\frac{\Gamma}{m l} + 
\frac{3C}{8l^3}\left[3Ck + \sqrt{3 k \left( 3 C^2 k + 16 \Gamma \frac{l^2}{m} \right)}\right]
\label{fapprox_edge}
\eeq
and:
\beq
f_a^{middle_{l<<\tilde{m}}}= \frac{3}{2}\frac{\Gamma}{\tilde{m} l} + 
\frac{3C}{8l^3}\left[3Ck + \sqrt{3 k \left(3 C^2 k + 8 \Gamma \frac{l^2}{\tilde{m}}\right)}\right]
\label{fapprox_middle}
\eeq

\begin{figure}
  \subfloat[$\tilde m=50\times 10^{-10}$meters]{\label{fig_short_width}\includegraphics[width=0.5\textwidth]{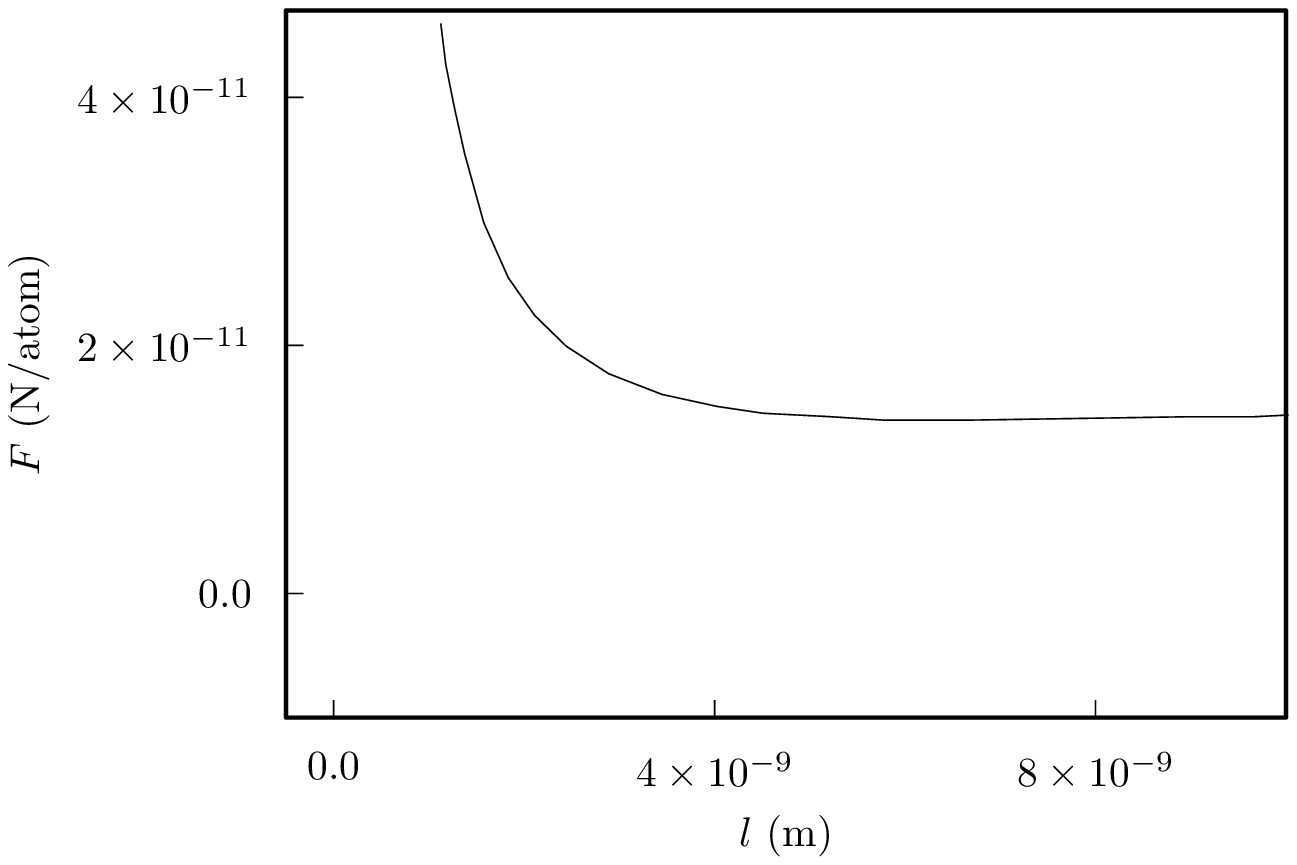}}
  \subfloat[$\tilde m=500\times 10^{-10}$meters]{\label{fig_long_width}\includegraphics[width=0.5\textwidth]{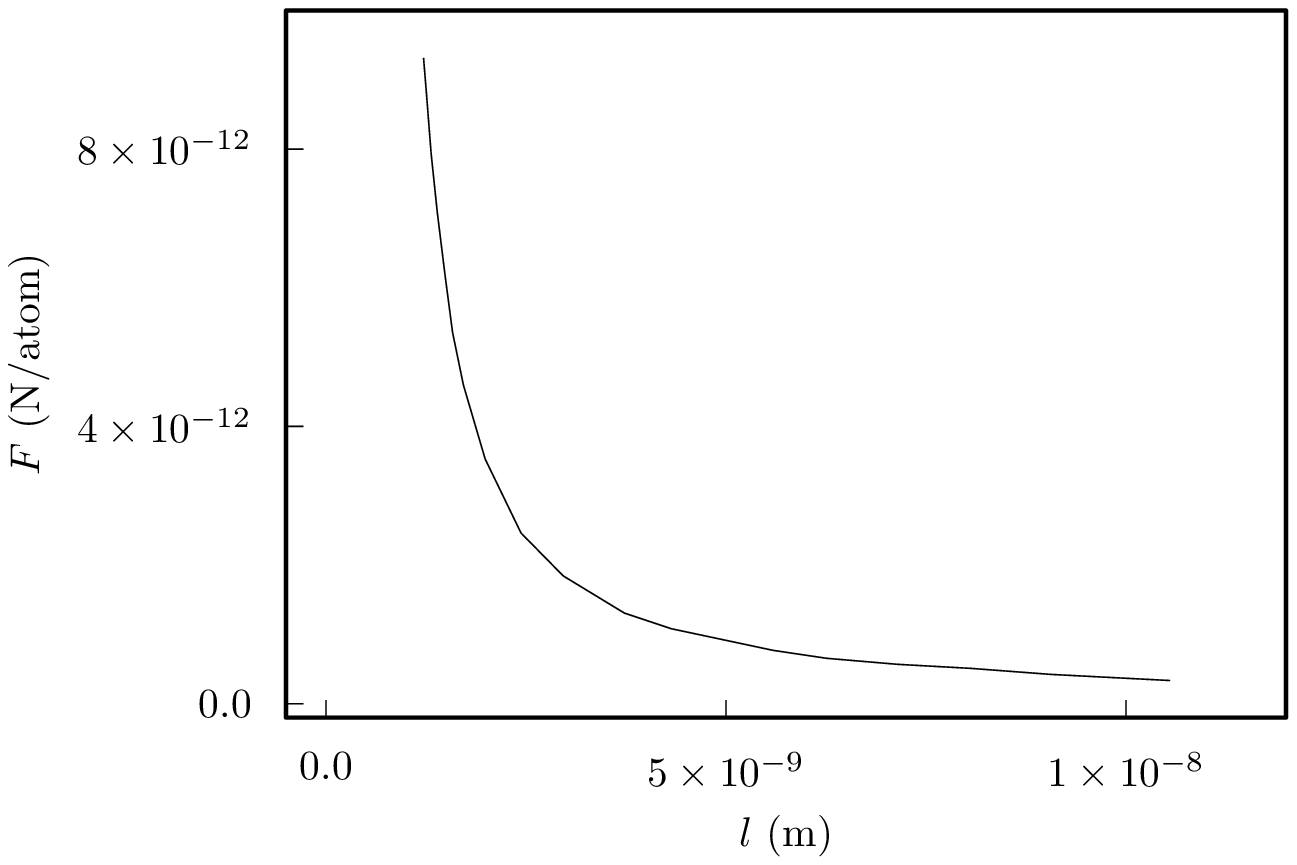}}
  \caption{Plot of force (in Newtons per atom) versus initial crack length (in meters), Eq.~\eqref{fgeneral_middle}. 
The half width of the sheet, $\tilde{m}$, in Fig.~\ref{fig_long_width} is 10 times larger than the one in Fig.~\ref{fig_short_width}. 
Note how the minimum force decreases with increasing initial crack length {\bf and} width of the sheets.}
  \label{fig_short_vs_long_width}
\end{figure}

It is intuitive 
that the force required for a crack to run will depend on the initial crack length.
For sheets of paper, for example, it is easier to tear a sheet
with a long crack, than to tear one with a short crack. 
Note that the expressions for the minimum force required for a crack to run, 
Eq.~\eqref{fgeneral_edge} and (\ref{fgeneral_middle}),
depend not only on the crack length, $l$,
but also on the width of the sheet, $m$, (or the half-width, $\tilde{m}$). 
Fig.~\ref{fig_short_vs_long_width} shows two graphs of force versus initial crack length
for a sheet with a crack in the middle. 
The half width of the sheet, $\tilde{m}$, in Fig.~\ref{fig_long_width} is 10 times larger than the one in Fig.~\ref{fig_short_width}.
Notice that longer initial crack lengths lead to lower minimum forces. Also, wider sheets lead to lower minimum forces.  
As a result, the minimum force approaches zero with increasing initial crack length {\bf and} width of the sheets. 

%

\section{Comparison between numerical and analytical results}
\label{sec_comparison}

To compare the numerical results with the analytical expression, Eqs.~(\ref{fgeneral_edge},\ref{fgeneral_middle}), we need the values 
of the bending modulus $k$ and the fracture toughness $\Gamma$ for graphene.

The experimental value of the bending modulus of graphene is $k_{exp}
= 1.2\textrm{eV}$ \cite{nicklow_lattice_1972}.  In previous numerical
work on ripples in graphene we found $k =1.77$eV
\cite{thompson-flagg_rippling_2009}.  Another numerical study, by
Fasolino {\it et al.} \cite{fasolino_intrinsic_2007}, obtained
$k=1.1$eV.

We have not been able to find an experimental measurement of graphene's
fracture toughness in the literature.  From our simulations we
obtained:
\beq
\Gamma_{numerical} \approx 3.82 \times 10^{-9}\textrm{J/m}
\label{gamma_value}
\eeq
For the full numerical calculation of graphene's fracture toughness see
Appendix~\ref{sec_gamma}.

The uniform downward force is applied to every atom on the sheet, therefore numerically we use force per atom $f_{atom}$, 
and not force per area $f_a$, as in the analytical calculations. 
The relationship between force per atom and force per area is:
\beq
f_{atom} = \frac{f_{a}} {\eta} = \frac{f_{a}} {38.17 \times 10^{18}\textrm{m}^{-2}} 
\label{fatom}
\eeq
where  $\eta = 38.17 \times 10^{18}\textrm{m}^{-2}$ is the number of atoms per area in graphene.

\subsection{Numerical study of a sheet with a crack in the middle} 

\begin{figure}
\includegraphics[width=.6\columnwidth]{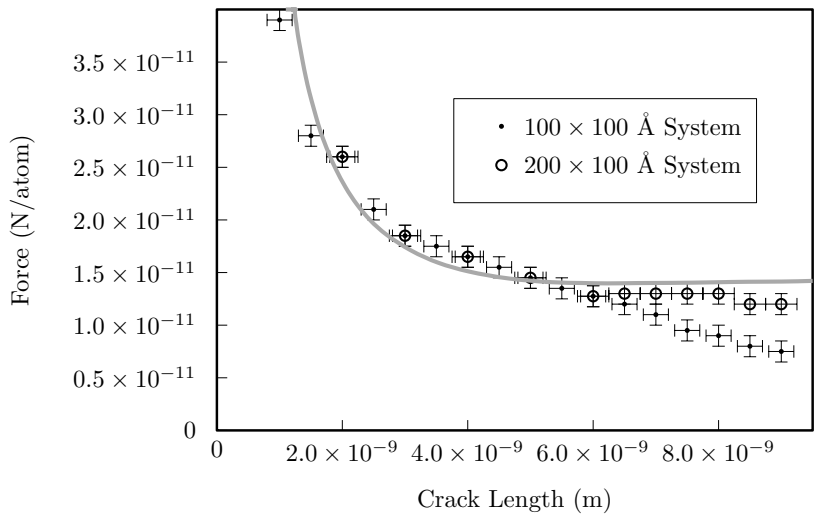}
\caption{Graph of force (in Newtons per atom) vs initial crack length (in meters) 
for a crack in the middle of the sheet. The line is the theoretical expression, Eq.~(\ref{fgeneral_middle}), 
the dots are the numerical results for a sheet of 100{\AA} by 100{\AA}, and the open circles are the numerical results for a sheet of 
200{\AA} by 100{\AA}.
}
\label{fig_100and200creasefit}
\end{figure}

A graph of force per atom versus initial crack length for a crack in
the middle of the sheet is shown in Fig.~\ref{fig_100and200creasefit}
The line is the theoretical expression, Eq.~(\ref{fgeneral_middle}), 
and the dots are the numerical results.
The horizontal error bars are estimated uncertainties of the initial
crack length due to the fact that the crack tip is not perfectly well defined.
The vertical error bars reflect the precision of our numerical simulations.

\begin{figure}
\includegraphics[width=.6\columnwidth]{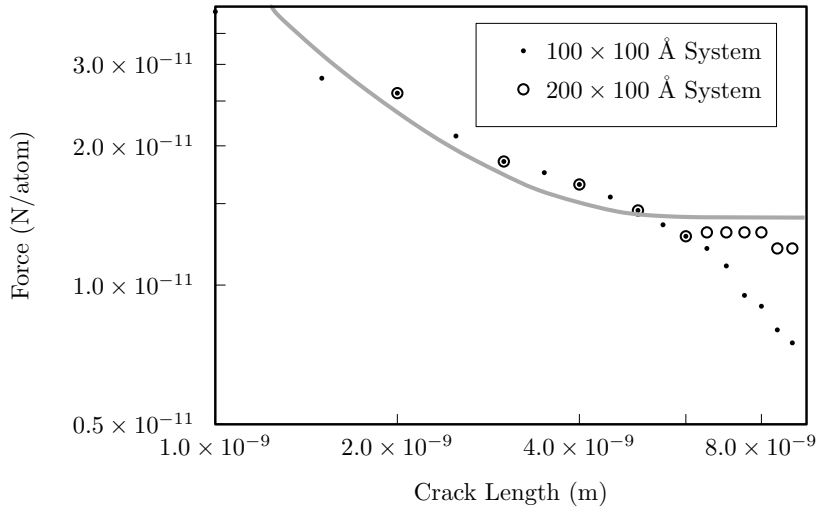}
\caption{Log-log graph of force (in Newtons per atom) vs initial crack length (in meters) 
for a wide-open middle crack. The line is the theoretical expression, Eq.~(\ref{fgeneral_middle}), 
the dots are the numerical results for a sheet of 100{\AA} by 100{\AA}, and the open circles are the numerical results for a sheet of 
200{\AA} by 100{\AA}.}
\label{fig_100and200creasefitloglog}
\end{figure}

Fig.~\ref{fig_100and200creasefit} shows good agreement between the theory and the simulations,
except for some of the longer cracks in systems of overall size
100{\AA} by 100{\AA}.  Discrepancies between theory and numerics
should be expected as crack lengths approach system dimensions. To
verify that system size effects were responsible we simulated the same
initial crack lengths in a sheet of 200{\AA} by 100{\AA}; that is, we
kept the same width, $m$, and changed only the sheet's length, a
parameter that does not appear in the theory. The results are shown in
Figs.~\ref{fig_100and200creasefit} and
\ref{fig_100and200creasefitloglog}.  The line is the theoretical
expression, Eq.~(\ref{fgeneral_middle}), the dots are the numerical
results for a sheet of 100{\AA} by 100{\AA}, and the open circle
are the numerical results for a sheet of 200{\AA} by
100{\AA}. Notice that for short initial cracks both results fall
exactly on top of each other. For longer initial cracks the force
plateaus for the 200{\AA} by 100{\AA} sheet, as in the theory,
while for the 100{\AA} by 100{\AA} sheet the force decreases due
to edge effects.


Another issue is that it is hard to produce short initial cracks in the middle of the sheet. They close easily and the uncertainty
of the crack tip plays an important role. Therefore, the uncertainties for the first two points on the graphs 
(initial cracks of 10{\AA} and 15{\AA}) perhaps should be higher than the
values presented on Fig.~\ref{fig_100and200creasefit}

\subsection{Numerical study of sheet with an edge crack} 

\begin{figure}
\includegraphics[width=.5\columnwidth]{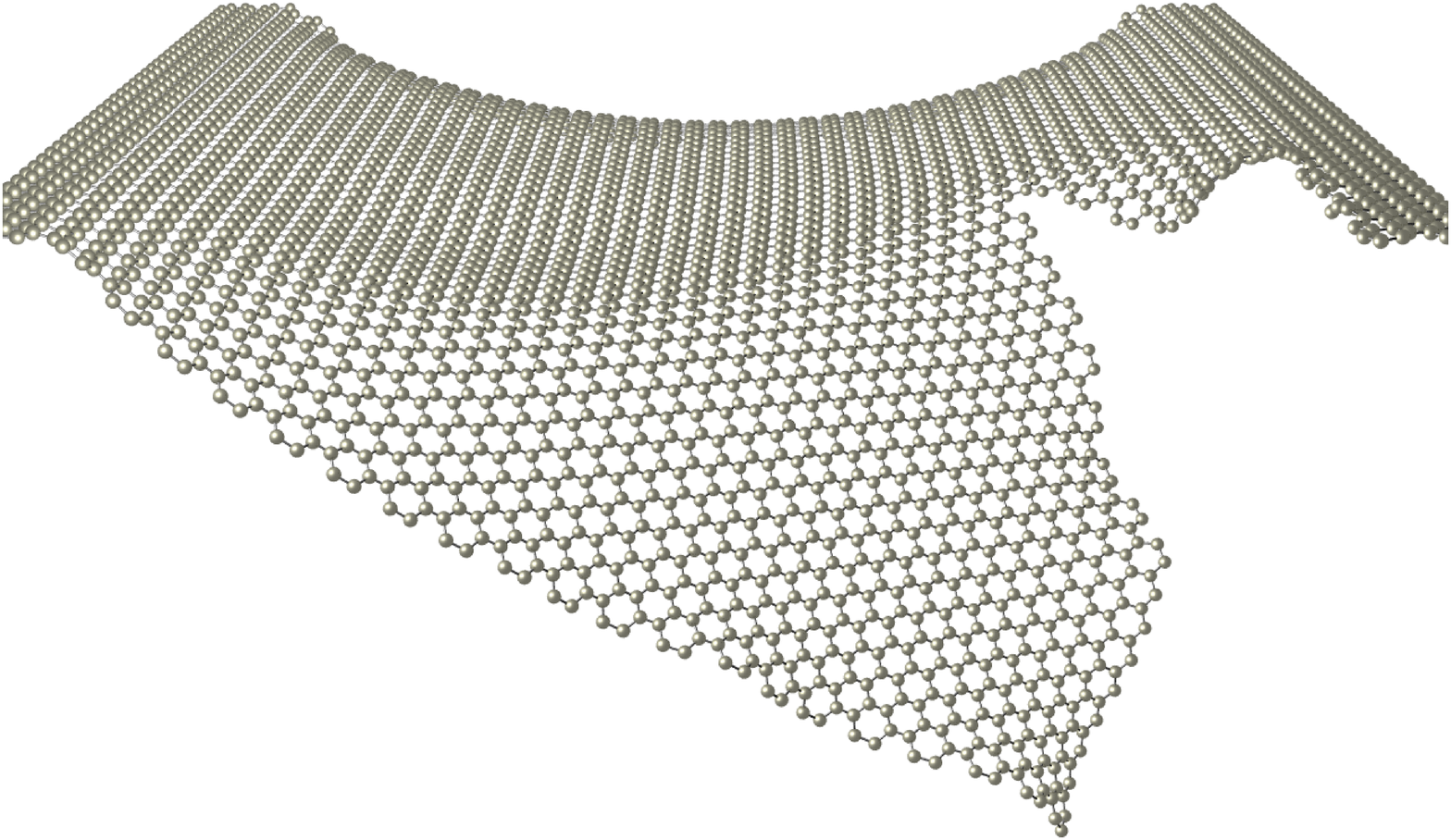}
\caption{Non-straight crack propagation in a 100{\AA} by 100{\AA} graphene sheet with an initial crack of $l \approx$ 25{\AA}. 
Note that the initial zigzag crack propagates as armchair.}
\label{fig_side_crack_not_straight}
\end{figure}

\begin{figure}
\includegraphics[width=.5\columnwidth]{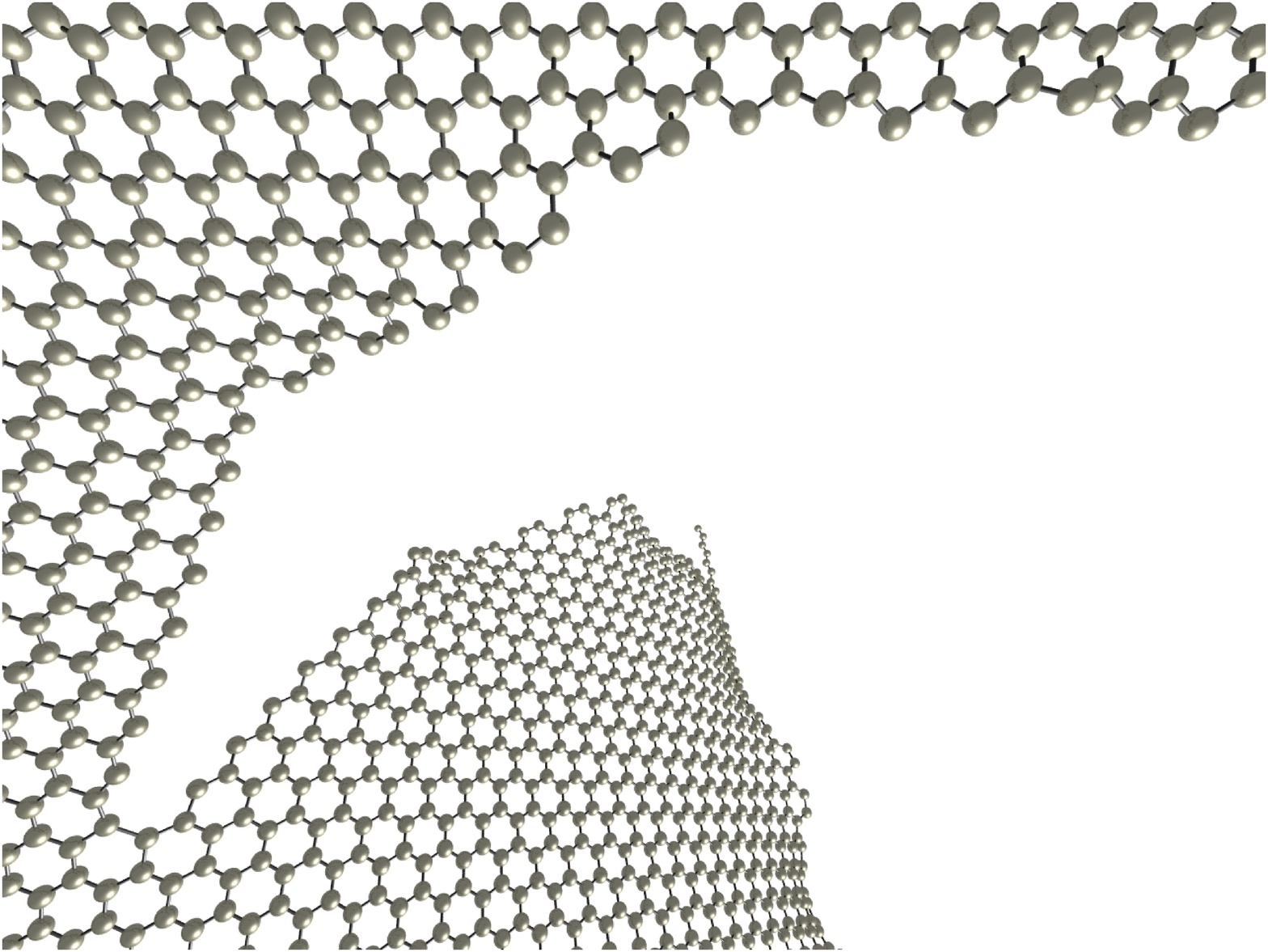}
\caption{Non-straight crack propagation in a 100{\AA} by 100{\AA} graphene sheet with an initial crack of $l \approx$ 30{\AA}. 
Note that the initial zigzag crack propagates along the 'armchair'
direction and then turns along the 'zigzag' direction.} 
\label{fig_side_crack_not_straight_top}
\end{figure}

Cracks at the edge of the sheet present some interesting characteristics, not seen in cracks in the middle of the sheet. 
The simulations show that, depending on the initial condition, a crack will not run straight through the sheet, 
as initially expected (see Fig.~\ref{fig_side_crack_not_straight} and Fig.~\ref{fig_side_crack_not_straight_top}). 
Short initial cracks run straight, while longer cracks do not. 

Another interesting result is that, depending on the initial crack length and orientation (zigzag or armchair), 
the propagation pattern will be different (see Fig.~\ref{fig_side_crack_not_straight} and Fig.~\ref{fig_side_crack_not_straight_top}). 
Similar dynamics have been seen in simulations of tearing graphene nanoribbons \cite{kawai_self-redirection_2009}.

\begin{figure}
\includegraphics[]{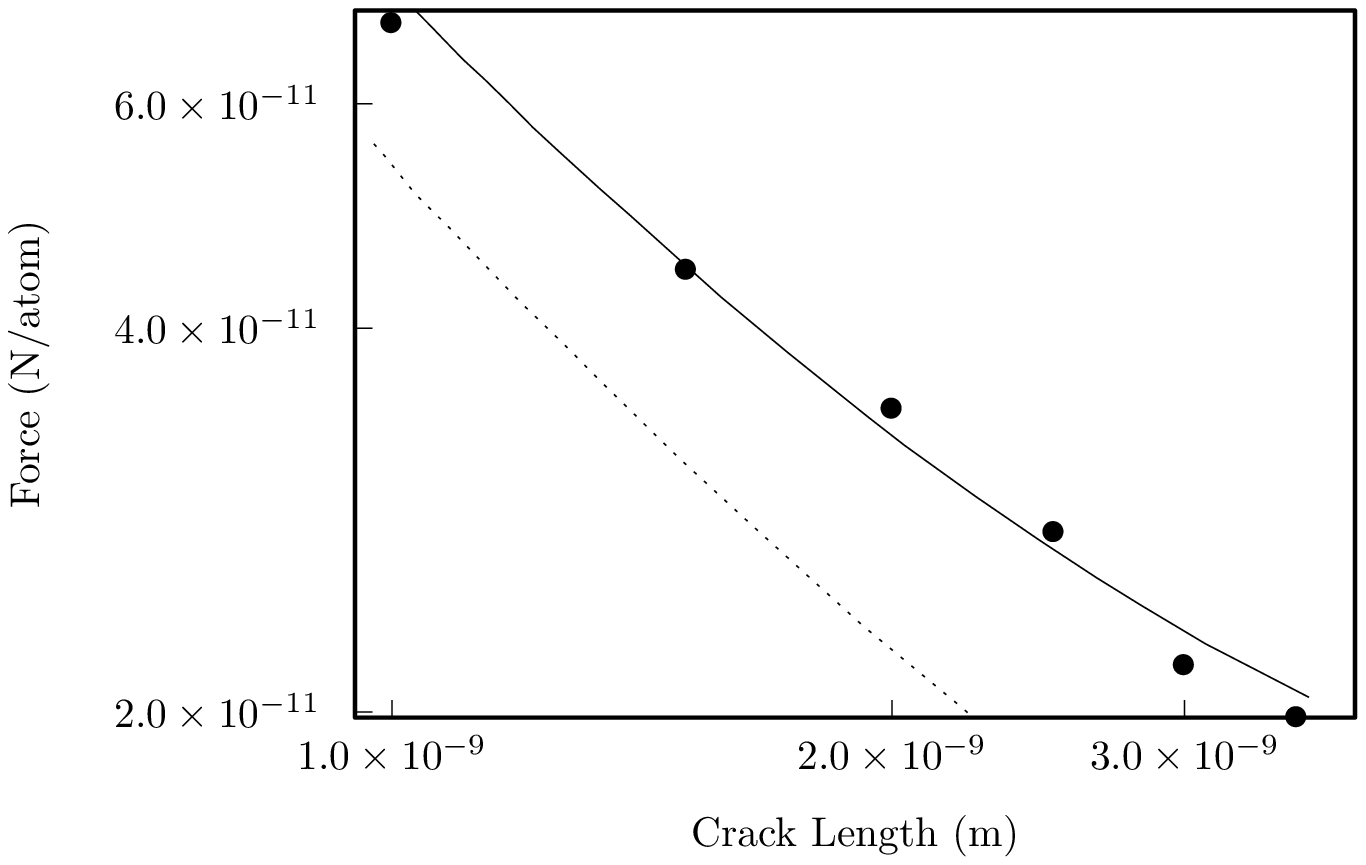}
\caption{Log-log graph of force  vs initial crack length 
for an edge crack. The dotted line shows Eq.~(\ref{fgeneral_edge}),
taking $m$ to be the width of the sheet minus the clamped region; the
solid line is the best fit for theoretical
expression, Eq.~(\ref{fgeneral_edge}), assuming $m$ to be a fitting
parameter
and the disks are  numerical results. 
The width of the sheet minus the clamped region is 90{\AA} and the best fit value for $m$ is 64{\AA}.}
\label{fig_logsidecreasebestfit}
\end{figure}

Fig.~\ref{fig_logsidecreasebestfit} shows a log-log graph of force per
atom versus initial crack length.  The dotted line is the theoretical
expression, Eq.~(\ref{fgeneral_edge}), and the disks are the numerical
results. Theory and numerical results do not agree very well.

\begin{figure}
  \subfloat[initial edge crack of $l \approx$ 10{\AA}]{\label{fig_side_crease_10A}\includegraphics[width=0.5\textwidth]{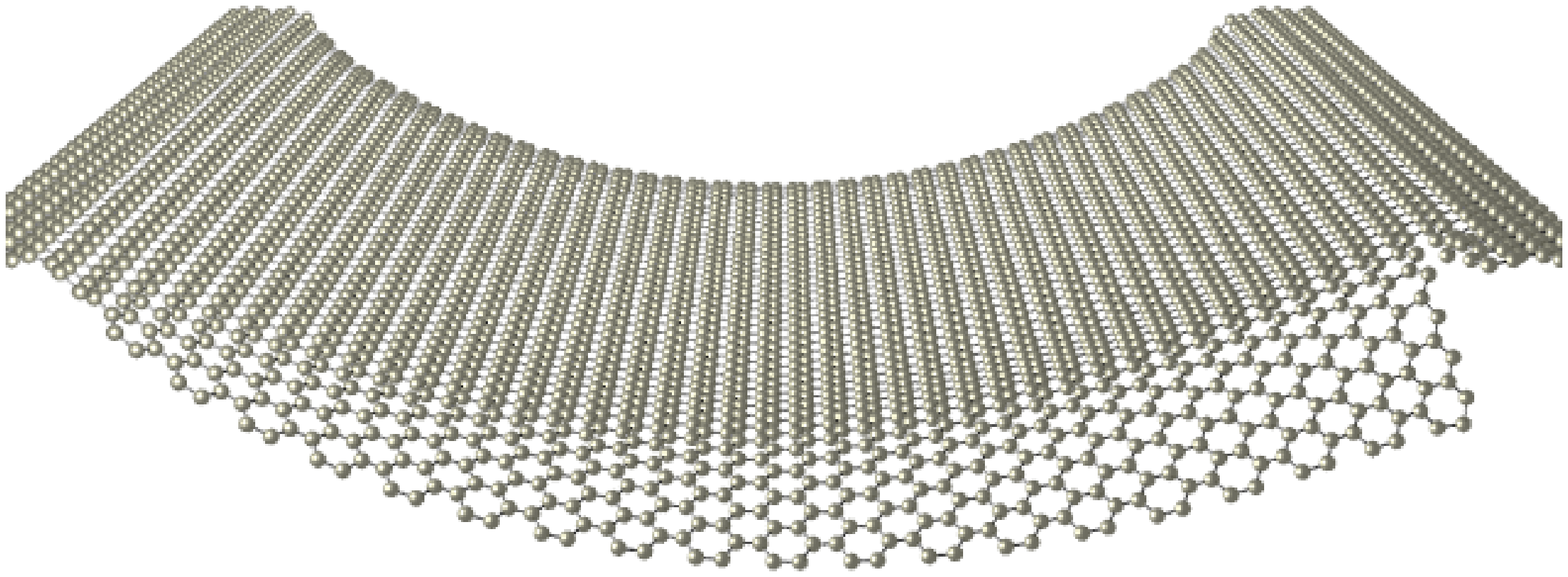}}
  \subfloat[initial edge crack of $l \approx$ 25{\AA}]{\label{fig_side_crease_25A}\includegraphics[width=0.5\textwidth]{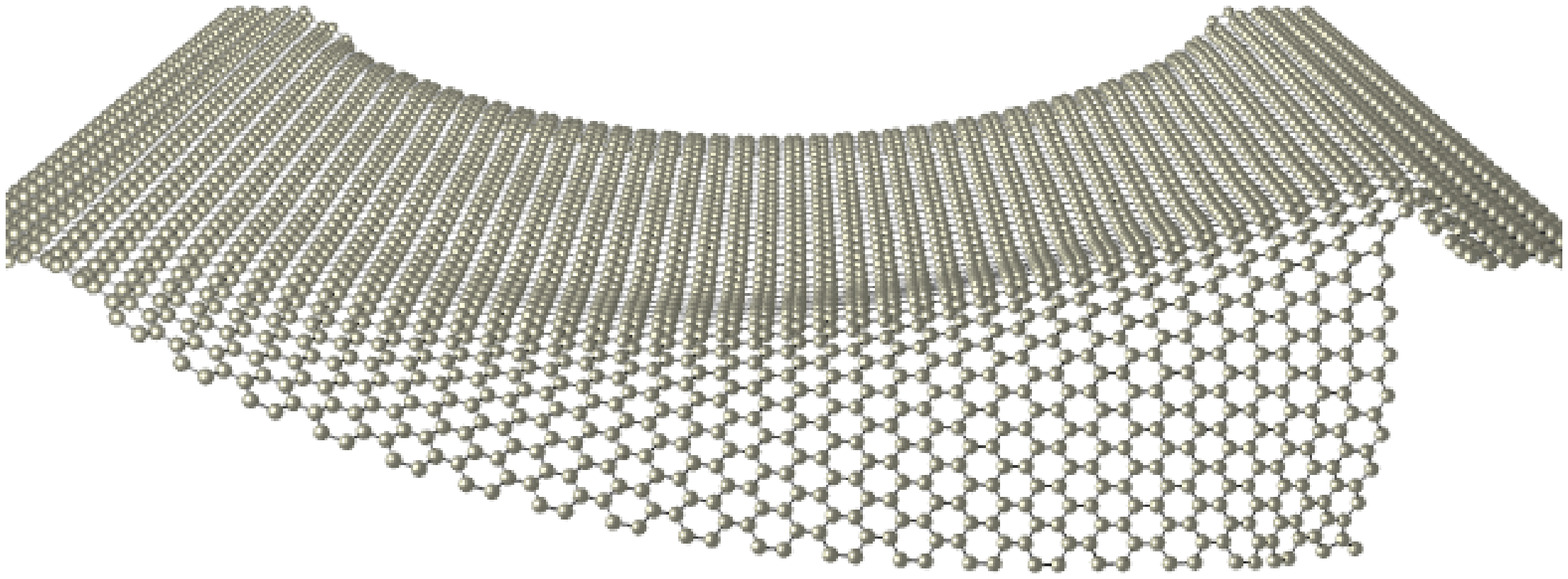}}
  \caption{Clamped, free-standing graphene sheets with an initial edge crack under a downward constant force.
The sheets exhibits a crease before the crack runs. 
The crease starts at the crack tip and ends before the fixed end.}
  \label{fig_multiple_side_crease}
\end{figure}

Looking at the simulations of sheets with an initial edge crack we
notice that the crease seems to end before the fixed end: see
Fig.~\ref{fig_multiple_side_crease}. This means that the value for $m$
should be smaller than the width of the sheet 
(minus the width of both fixed ends). The best fit for the theoretical expression
Eq.~(\ref{fgeneral_edge}) with the numerical results, assuming $m$ to
be a fitting parameter, is shown as the solid line in
Fig.~\ref{fig_logsidecreasebestfit}. The width of the sheet minus the
clamped region is $\approx$ 90{\AA} and the best fit value for $m$ is
64{\AA}. It was difficult to determine the crease length from
the simulations, so we treated it as a fitting parameter, obtaining
very reasonable results.

%

\section{Conclusion}
\label{sec_conclusion}

We presented a numerical and an analytical study of the propagation of cracks in clamped, free-standing graphene as a function of the out-of-plane force. 
The geometry is motivated by experimental configurations that expose graphene sheets to out-of-plane forces, such as the back-gate voltage. 
We studied two different initial conditions: a crack at the edge of the sheet and a crack in the middle of it.

We obtained approximate analytical expressions for the minimum force required for a crack to run. These expressions depend on the initial crack length, as expected, but also on the width of the graphene sample. 
The minimum force decreases with increasing initial crack length and  increasing width of the sheets.

The numerical fracture forces show good agreement with the analytical fracture forces for cracks in the middle of the sheet.
For cracks at the edge the numerical results and the theory do not
agree as well, unless we consider crease configurations that do not
reach all the way to the edge of the sample.  
The simulations show that depending on the initial condition a crack will not run straight through the sheet, as initially expected. Initial cracks in the middle of the sheet always run straight, while initial cracks at the edge do not.

\begin{figure}
\includegraphics[width=.5\columnwidth]{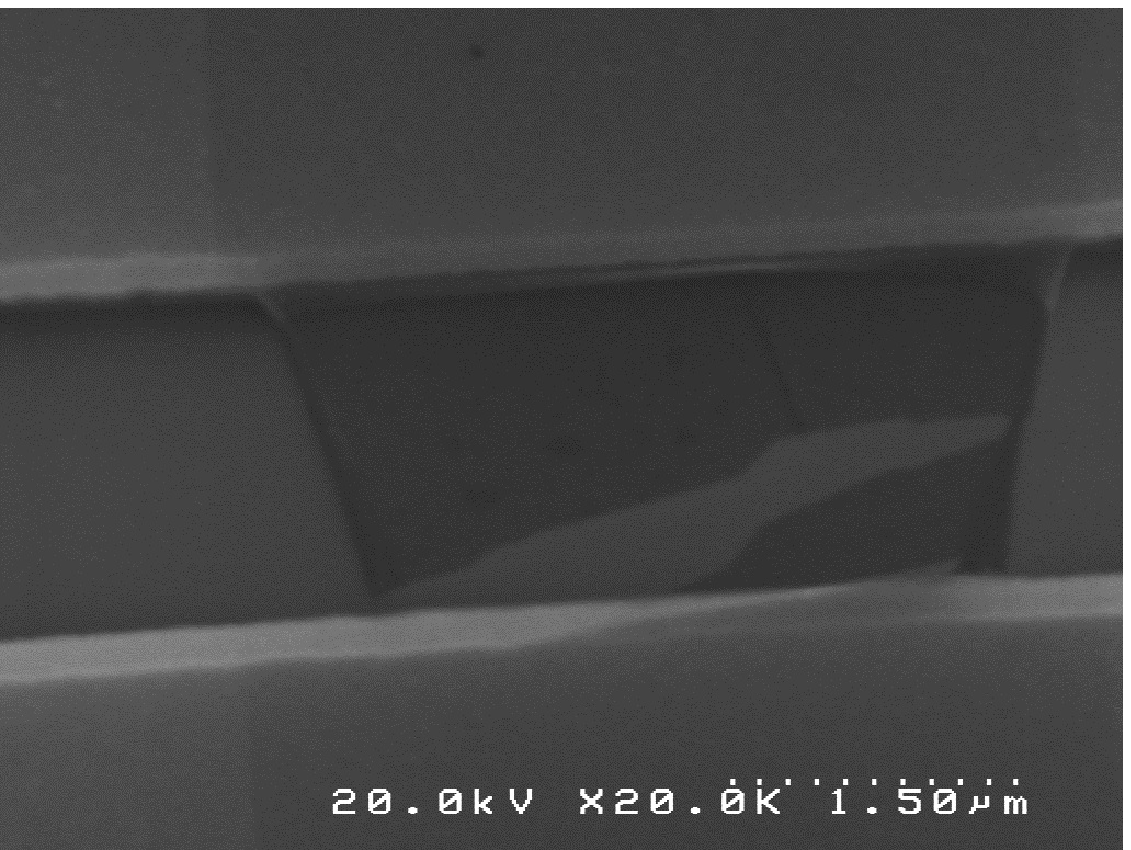}
\caption{Top view: SEM image of free-standing CVD graphene. Note the crack runs from the edge to the middle of the sheet, making sharp turns, similar to our simulations. Image courtesy of the Bolotin Research Group.}
\label{fig_exp_topview_crack}
\end{figure}

Depending on the initial crack orientation (zigzag or armchair), the
propagation pattern will be different. Similar dynamics have been seen
in simulations of tearing graphene nanoribbons
\cite{kawai_self-redirection_2009} and in experiments, see
Fig.~\ref{fig_exp_topview_crack}.  The edge orientation of a graphene
sheet determines its electronic properties, therefore it will be most
useful to be able to predict the edge orientation of produced samples.

\begin{figure}
\includegraphics[width=.5\columnwidth]{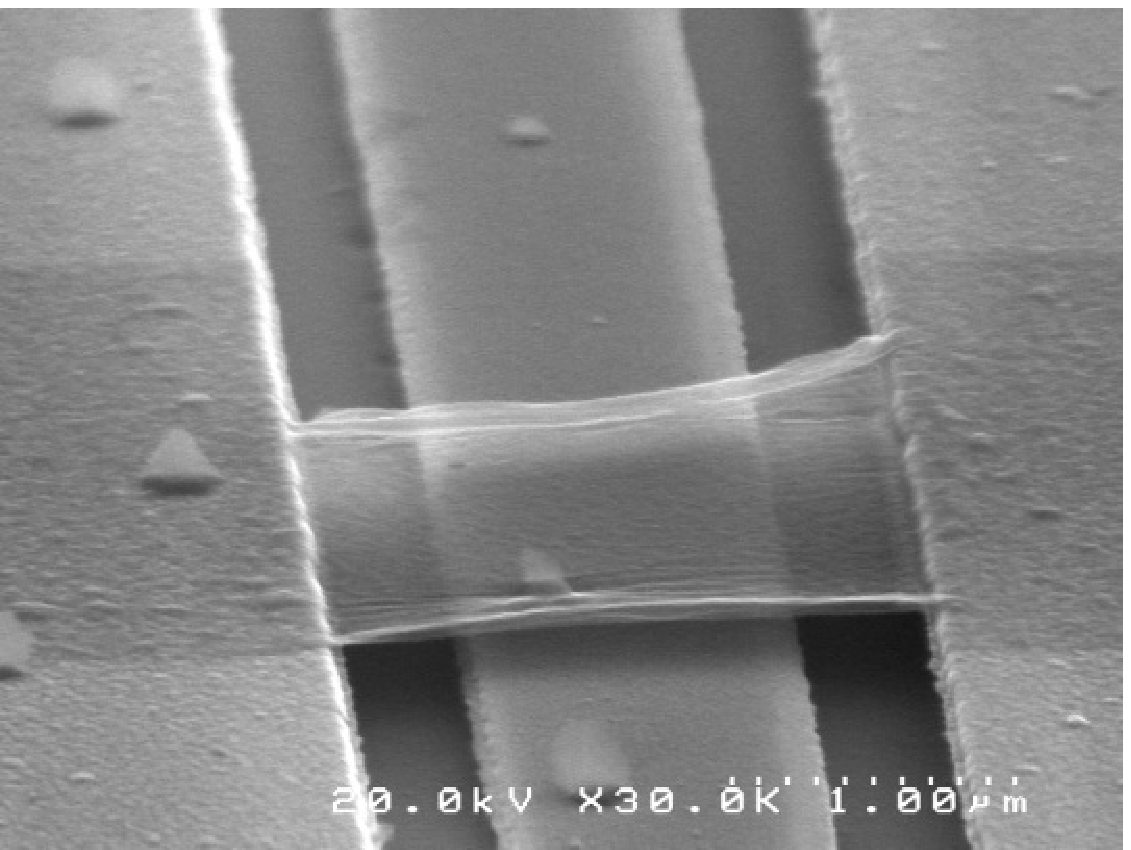}
\caption{SEM image of free-standing CVD graphene. Note the crack at the left edge and the crease that runs all the way to the right edge, similar to what we observe in our simulations. Image courtesy of the Bolotin Research Group.}
\label{fig_exp_crease}
\end{figure}

Another interesting numerical result is that the sheet folds in a
crease before the crack runs, both for sheets with a crack in the
middle (on both sides of the crack) and at the edge,
Fig.~\ref{fig_crease}.  In experiments, folds, scrolls, and creases
are commonly observed in free-standing graphene samples, see
Fig.~\ref{fig_exp_crease}.


It has been difficult to gather experimental data on this problem,
because the fracture of free-standing graphene sheets is in general an
undesirable occurrence.  Therefore most experiments do not report or
take measurements of such events.  Preliminary results \cite{Bolotin}
show that, for initial crack lengths of about 10\% of the graphene
sample's length, our expression for the minimum force required for a
crack to run, Eq.~\eqref{fgeneral_edge} and \eqref{fgeneral_middle},
results in forces comparable to the forces a free-standing graphene is
subjected to in a back-gate voltage experiment.

In particular, for samples of roughly 2$\mu$m by $2\mu$m, fracture
was sometimes observed with back-gate voltages of 2-3 V with the
sample at a  height of  150 nm above the substrate. We estimate
an electric field of $20\times 10^6$ V/m, leading to a force per atom
of $10^{-16} N$/atom. According to our calculations, this force is too
small to cause fracture for seed cracks of reasonable length. For this
value of the force per atom, we would need a seed crack of length
$1.5\mu$m. Alternatively, for seed cracks of length $0.2\mu$m, we
would not expect fracture until the force reached $8\times10^{-16}
N$/atom. While this discrepancy is admittedly large, we note that we
are working without an experimental value for fracture toughness, and
that classical potentials can easily lead to inaccurate values for this
quantity.

The analytical expressions for the the minimum force required to tear
a two dimensional sheet, such as graphene, in terms of the initial
crack length, offer insight into the tearing of graphene, and suggest
the order of magnitude for the forces to be used or avoided in
experiments.  Also, experiments can obtain the value for the tearing fracture toughness of graphene (a quantity we have not been able
to find in the literature) by combining our theory with measurements
of initial crack length and applied force.

%
%

\appendix
\section{ Energy for bending a 1D strip}
\label{sec_full_1Dbending}

The energy of a 1D strip of length $L$ under a downward force $f_l$
is given by (Eqs.~(\ref{U1Dbend}) and (\ref{no_small_bend})):
\begin{equation}
U^{1D} = \int_{0}^{L} ds \left[\frac{k_l}{2} \left(\frac{d \theta}{ds}\right)^2 + f_{l} \int_{0}^{s} \sin(\theta(s')) ds'\right]
\label{U1D_no_small_bend_again}
\end{equation}

Note that $s'$ is just a variable ({\bf not} a derivative of $s$).

We need to minimize Eq.~(\ref{U1D_no_small_bend}) to obtain the energy outside the crack tip in terms of the minimum 
force required for the crack to run.
To find the minimum (or the maximum) of $U^{1D}$ we take the functional derivative and make it equal to zero:
\begin{equation}
\lim_{\epsilon->0} \frac{U^{1D}(\theta+\epsilon\phi) - U^{1D}(\theta)}{\epsilon}=0 
\end{equation}

This results in:
\begin{equation}
\lim_{\epsilon->0} \frac{1}{\epsilon} \int_{0}^{L} ds \left\{ \frac{k_l}{2} \left[2\epsilon\frac{d\phi}{ds}\frac{d\theta}{ds}+
\epsilon^2 \left( \frac{d\phi}{ds} \right)^2 \right] + f_{l}  \int_{0}^{s} \left[\sin(\theta(s')+\epsilon\phi(s')) - 
\sin(\theta(s'))\right] ds'\right\}=0 
\end{equation}

Expanding the sine function in a Taylor series we obtain:
\begin{equation}
\lim_{\epsilon->0} \frac{1}{\epsilon} \int_{0}^{L} ds \left\{\frac{k_l}{2} \left[2\epsilon\frac{d\phi}{ds}\frac{d\theta}{ds}+
\epsilon^2 \left( \frac{d\phi}{ds} \right)^2 \right] + f_{l} \int_{0}^{s} \left[\epsilon \phi(s') \cos(\theta(s')) + O(\epsilon^2)\right] 
ds'\right\}=0 
\end{equation}

Dividing by $\epsilon$ and taking the limit as it goes to zero:
\begin{equation}
\int_{0}^{L} ds \left\{\frac{k_l}{2} 2 \frac{d\phi}{ds}\frac{d\theta}{ds} + f_{l} \int_{0}^{s} \phi(s') \cos(\theta(s')) ds'\right\}=0 
\end{equation}

Using integration by parts on the first term and the fact that $\phi$ vanishes at 0 and L:
\begin{equation}
\int_{0}^{L} ds \left\{- k_l \phi\frac{d^2\theta}{ds^2} + f_{l} \int_{0}^{s} \phi(s') \cos(\theta(s')) ds'\right\}=0 
\end{equation}

The result should be independent of the test function $\phi$. Here we choose $\phi=\delta(s-s'')$: 
\begin{equation}
\int_{0}^{L} ds \left\{- k_l \delta(s-s'') \frac{d^2\theta}{ds^2} + f_{l} \int_{0}^{s} \delta(s'-s'') \cos(\theta(s')) ds'\right\}=0 
\end{equation}

After some algebra we obtain:
\begin{equation}
- k_l \frac{d^2\theta(s'')}{ds''^2} + f_{l} \cos(\theta(s'')) (L - s'') =0
\label{diff_eq_1D_no_small} 
\end{equation}

Note that $s''$ is just a variable ({\bf not} the second derivative of $s$).
At this point an appropriate approximation needs to be considered in order to solve this equation analytically.

\subsubsection{One-dimensional crease energy}

Here we consider the limit $L \rightarrow \infty$, since by the time $s''$ is comparable to $L$, $\theta=-\pi/2$.
Eq.~(\ref{diff_eq_1D_no_small}) then simplifies to:
%
%
\begin{equation}
- k_l \frac{d^2\theta(s'')}{ds''^2} + f_{l} \cos(\theta(s''))L  =0 
\end{equation}

After some manipulation we obtain:
\begin{equation}
\frac{d\theta(s'')}{ds''} = \sqrt{2\frac{f_{l}L}{k_l} \sin(\theta(s'')) - C_1} 
\label{dthetatds}
\end{equation}
where $C_1$ is a constant, and it can be determined from the following conditions:
\begin{equation*}
     \frac{d\theta}{ds''} (L)=0   
      \end{equation*}

 \begin{equation*}
       s''\rightarrow L, \theta \rightarrow -\frac{\pi}{2}
      \end{equation*}
The result is:
\begin{equation}
C_1 = 2\frac{f_{l}L}{k_l} \sin(\theta(L)) = 2\frac{f_{l}L}{k_l} \sin\left(-\frac{\pi}{2}\right) =  - 2\frac{f_{l}L}{k_l}
\end{equation}

Inserting into Eq.~(\ref{dthetatds}) one obtains
\begin{equation}
\frac{d\theta(s'')}{ds''} = \sqrt{2\frac{f_{l}L}{k_l} \left[\sin(\theta(s'')) +1\right]} .
\end{equation}
%


Substituting
\begin{equation}
 \theta(s'')=\frac{\pi}{2} -\Phi(s'') 
\label{thetatophi}
\end{equation}
and using trigonometric identities, we obtain:
\begin{equation}
\frac{d\Phi(s'')}{ds''} = - 2\sqrt{\frac{f_{l}L}{k_l}}\cos(\Phi(s'')/2) 
\end{equation}

After some manipulation we have 
\begin{equation} 
\Phi(s'') = 2 \arccos \left( \sech \left(\sqrt{\frac{f_{l}L}{k_l}} s'' - C_2 \right) \right)
\end{equation}
where $C_2$ is a constant to be determined.

Using Eq.~(\ref{thetatophi}) we can go back to $\theta$:
\begin{equation} 
\theta(s'') = \frac{\pi}{2} - 2 \arccos \left( \sech \left(\sqrt{\frac{f_{l}L}{k_l}} s'' - C_2 \right) \right)
\end{equation}

Applying the condition that $\theta(0)=0$, we obtain the expression for the bending function of a 1D strip
\begin{equation} 
\theta(s'') = \frac{\pi}{2} - 2 \arccos \left( \sech \left(\sqrt{\frac{f_{l}L}{k_l}} s'' + \arccosh\left(\sqrt{2}\right) \right) \right).
\label{theta_no_small_bend}
\end{equation}

Now that we have $\theta$ we insert it back in the expression for the energy of a 1D strip, Eq.~(\ref{U1D_no_small_bend_again}), and find
\begin{eqnarray}\
 U^{1D}=
-2 \sqrt{2} \sqrt{f_l k_l L} \nonumber \\
+ \frac{2 k_l}{L} \ln\left[\cosh\left(\sqrt{\frac{f_l L^3}{k_l}}\right) 
+ \frac{1}{\sqrt{2}}\sinh\left(\sqrt{\frac{f_l L^3}{k_l}}\right)\right] \nonumber \\
+2 \sqrt{f_l k_l L} \tanh\left(\sqrt{\frac{f_l L^3}{k_l}} + \arccosh\left(\sqrt{2}\right)\right)
 -\frac{1}{2}f_lL^2.
\label{U1D_crease}
\end{eqnarray}

Looking at the asymptotic behavior for large $L$
\beq
U^{1D}=  2\left(2- \sqrt{2}\right)\sqrt{f_l k_l L} - \frac{1}{2}f_lL^2
\eeq
we finally arrive to the result presented in Eq.~(\ref{U1D_crease_large_L}).

\section{Numerical calculation of graphene's fracture toughness $\Gamma$}
\label{sec_gamma}

The growth of a crack requires the creation of two new surfaces and hence an increase in the fracture toughness (i.e. surface energy). 
As graphene is a two-dimensional sheet the new surfaces are actually edges and  
the fracture toughness is given by energy per length, and not the usual energy per area.

\begin{figure}[h]
\includegraphics{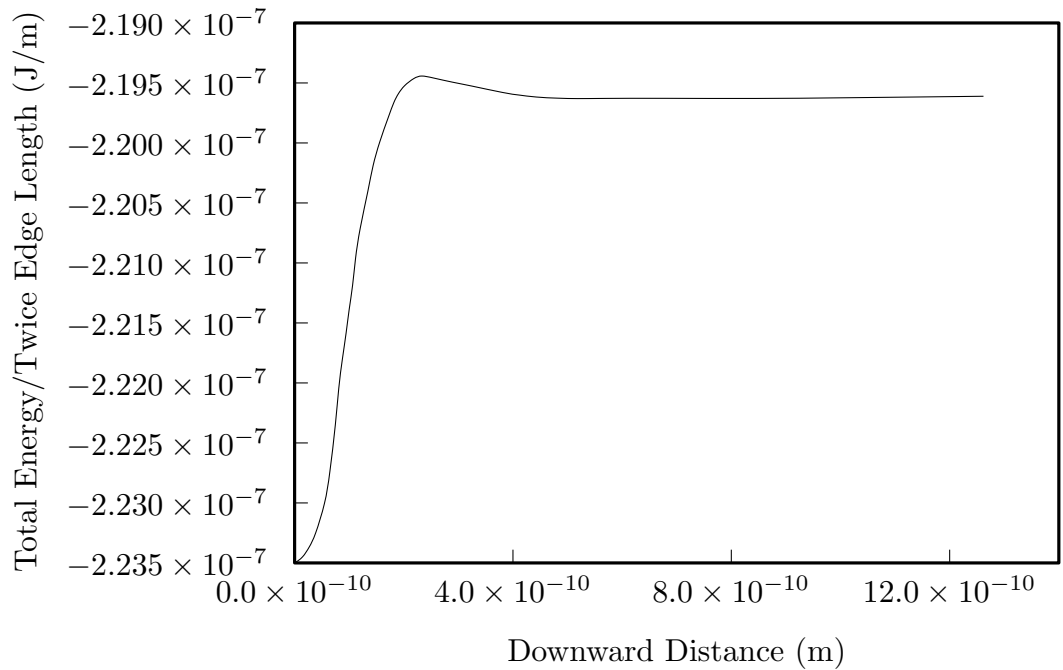}
\caption{Graph of energy per length (in Joules per meter) vs downward distance (in meters).}
\label{gamma_graph}
\end{figure}

Numerically we can find the energy to create a new surface by simply separating the sheet from its fixed edges. 
This energy only depends on the interaction between atoms, obtained from the MEAM potential. 
Therefore here we do not apply a downward force, we do not have initial cracks, and no molecular dynamics is done. 
We start by fixing two sides of a graphene sheet. Then we move the rest of the sheet downward. Every time step the atoms are moved down 
by the same distance. The sheet moves farther and farther away from its fixed ends until finally the new surfaces are created, 
Fig.~\ref{gamma_graph}. 
The fracture toughness $\Gamma$ is then given by the change in energy divided by the length of the two edges created: 
\begin{equation}
 \Gamma_{numerical} =\frac{(E_{final}-E_{initial})}{\textrm{2 $\times$ edge length}}
 \approx 3.82 \times 10^{-9}\textrm{J/m}.
\label{gamma_value_again}
\end{equation}

This calculation does not substitute an experimental measurement of the fracture toughness.
It is fundamental to the theory of fracture that the fracture toughness be measured experimentally. 
We have not been able to find experimental values for this quantity in the literature. 

%

\section*{Acknowledgement}

We thank Allan MacDonald for pointing out this problem to us, Hiram Conley and Kirill Bolotin for experimental information and images, Fulbright and CAPES for scholarship funding, 
and the National Science Foundation for funding through DMR 0701373.

\bibliographystyle{apsrev}
\bibliography{bibliography}

\end{document}